\documentstyle[epsf,epsfig,12pt]{article}  
\begin{document}

\newcommand{\be}{\begin{equation}}

\begin{titlepage}

\pagenumbering{arabic}
\vspace*{-1.cm}
\begin{tabular*}{15.cm}{l@{\extracolsep{\fill}}r}
&
\hfill \bf{DEMO-HEP 97/08}

\\
& 
\hfill 20 November, 1997
\\
\end{tabular*}
\vspace*{2.cm}
\begin{center}
\Large 
{\bf A Method to include Detector Effects in Estimators sensitive to the 
Trilinear Gauge Couplings }
\\
\vspace*{2.cm}
\normalsize { 
{\bf G. K. Fanourakis, D. Fassouliotis and S. E. Tzamarias} \\
\vspace*{0.3cm}
{\footnotesize N.C.S.R. Demokritos}\\
}
\end{center}
\vspace{\fill}
\begin{abstract}
\noindent
This paper describes the use of weighted Monte Carlo events to
accurately approximate integrals of functions of the
experimentally measured kinematical vectors and their
dependence on physical parameters. 
This technique is demonstrated in estimating
the evolution of cross sections, efficiencies, measured kinematical 
distributions and mean values as functions of the Trilinear Gauge Couplings. 

\end{abstract}
\vspace{\fill}

\end{titlepage}

\pagebreak

\begin{titlepage}
\mbox{}
\end{titlepage}

\pagebreak

\setcounter{page}{1}    


\section{Introduction}

 The accurate estimation of the ElectroWeak parameters
(in the following just parameters) at LEPII demands attention
to the definition of the probability density
functions (p.d.f.). Measurement effects
such as the detector resolution and the selection efficiency have
to be convolved with the physics functions in order to describe
the distribution of the observed kinematical variables.
Due to the fact that detector effects are not easily parametrized,
one usually employs Monte Carlo (M.C.)
integration techniques. The disadvantage of this method, though,
is that the integrated
quantities are in principle by themselves functions of the parameters
(e.g. the cross section or the efficiency corresponding to a part of the phase
space) and consequently
several M.C. sets of events must be used to cover the parametric space.

This paper describes a reweighting technique  which covers a wide range of
values
of the parameters (extrapolation area) whilst using a single M.C.
set of events.
Furthermore this technique can be extended by combining
several sets of M.C. events, produced with different parameter
values, and thus enlarging the extrapolation area in addition to minimizing
the statistical error. 

The general principles of reweighting are
demonstrated in calculating  the cross sections in  multidimentional bins of
the observed kinematical variables and their dependence on the parameter
values.
This particular example has important applications in the shape definition  of
the kinematical distributions which are used in measuring the Trilinear Gauge
Couplings (TGC's). Other estimating procedures compare the average of
certain functions of the observed kinematical vectors with the phenomenological
expectations convolved with the resolution and efficiency functions. It is
shown,
in this paper, that by employing the reweighting technique this convolution is
performed accurately
and a very efficient estimation of the TGC's  can be made.
This work finally concludes with a demonstration of the accuracy and consistency of the technique by numerical results.

\section{Reweighting a Single Set of M.C. events}

 Throughout this analysis the symbol V $={V_{1}, \ldots ,V_{k}}$ denotes the
k-dimensional kinematical vector which defines completely a M.C.  event at
generation (generation vector) whilst
$\Omega ={  \Omega_{1}, \ldots ,\Omega_{k}}$  stands for the reconstructed
equivalent of V (reconstructed or observed vector).
The following symbols are also defined:
\begin{description}
\item[{\bf $\bar{\alpha}$}] $= { \alpha_{1}, \ldots ,\alpha_{\rho } }$ denoting
the $\rho$ parameters which are required to define the p.d.f.
\item[{\bf $d \sigma (V, \bar{\alpha } )/dV$}] denoting  the differential
cross section as a function of the generated vector.
\item[{\bf $R(V,\Omega )$}] denoting the resolution function, i.e. the
probability of an
event generated with V to be observed with $\Omega$ kinematical vector.
Obviously \footnote{Where the limits of an integration are not explicitly
stated an integration all over the phase space is meant.}
\begin{equation}
\int R(V,\Omega )d \Omega = 1
\end{equation}
\item[{\bf $\epsilon (V)$}] denoting the probability of an event generated with
V to be
observed.
 \item[{\bf $d \tilde{\sigma }(\Omega , \bar{\alpha } ) /d \Omega $}]
denoting the differential observed
cross section as a function of the reconstructed vector which is defined as:
\begin{equation}
\frac{d \tilde{\sigma } (\Omega , \bar{\alpha } ) }{d \Omega } =
\int\frac{d \sigma (V, \bar{\alpha } ) }{dV}\cdot \epsilon (V)\cdot R(V,\Omega )dV
  \label{eq:a}
\end{equation}
\item[{\bf $\sigma_{obs}(\bar{\alpha })$}] denoting the total observed cross
section.
\item[{\bf $g(V,\bar{\alpha })$}] denoting the p.d.f. defined as:
\begin{equation}
g(V,\bar{\alpha }) = \frac{(d \sigma (V, \bar{\alpha })/dV)\cdot \epsilon (V)}
{\sigma_{obs}(\bar{\alpha })}  \label{eq:b}
\end{equation}
\end{description}
Following the above definitions the joint distribution of the observed
kinematical vectors is described by:
\begin{equation}
G(\Omega,\bar{\alpha }) = \int g(V,\bar{\alpha })\cdot R(V, \Omega )dV
\label{eq:ppp}
\end{equation}

\subsection{ The Number of Expected Events in a Part of the Phase Space.}

 Let $\Delta \Omega$ be a part of the observed phase space and let
the integral I to be the corresponding cross section
\begin{equation}
I = \int_{\Delta \Omega }\frac{d \tilde{ \sigma} (\Omega , \bar{\alpha})}{d
\Omega }
d\Omega  \label{eq:1}
\end{equation}
for  a set  $\bar{\alpha}$ of parameter values.
 The number of the expected events,
$\mu (\bar{\alpha } ,\Delta \Omega $), in the
$\Delta$$\Omega$ part of the phase space for an accumulated luminosity L is:
\begin{equation}
\mu (\bar{\alpha }, \Delta \Omega ) = I\cdot L  \label{eq:2}
\end{equation}

Let us  assume that there is a set of $N_{0}$ M.C.
events  generated with the p.d.f.
$g(V,\bar{\alpha _{g}})$  with $\bar{\alpha _{g}}$
parameter
values. Let also N be the number of the above events which have
survived the acceptance, reconstruction and selection criteria after
full detector simulation whilst $n_{\Delta \Omega}$ of them lie inside
$\Delta \Omega$.
By multiplying and dividing the integrant of (\ref{eq:1}) by the properly
normalized p.d.f.
\begin{equation}
\frac{g(V,\bar{\alpha _{g}})}{D(\Delta \Omega ,\bar{\alpha _{g}})}
\label{eq:2b}
\end{equation}
 the integral I is written as:
\begin{equation}
I = D(\Delta \Omega ,\bar{\alpha _{g}})\cdot\int_{ \Delta \Omega }
\int\frac{d\sigma (V, \bar{\alpha })}{dV}\cdot\frac{\epsilon (V)}
{g(V,\bar{\alpha _{g}} )}\cdot
\frac{g(V,\bar{\alpha _{g}} )
\cdot R(V, \Omega )}{D(\Delta \Omega ,\bar{\alpha _{g}})}
dVd\Omega  \label{eq:3}
\end{equation}
where
\begin{equation}
D(\Delta \Omega ,\bar{\alpha _{g}}) =
\int_{\Delta \Omega }\int g(V,\bar{\alpha _{g}} )
\cdot R(V,\Omega )dVd\Omega    \label{eq:4}
\end{equation}

Equation (\ref{eq:3}) can be written in a simpler form, by
expressing the differential
cross section $d \sigma (V, \bar{\alpha } )/dV$ as a product of the matrix
element squared ${\cal{M}}(V, \bar{\alpha } )$ and of
the Lorentz Invariant Phase Space dLIPS(V). Specifically:
\begin{equation}
 I = D(\Delta \Omega ,\bar{\alpha _{g}})
\cdot \sigma_{obs}(\bar{\alpha _{g}})\cdot <w(V,\bar{\alpha },\bar{\alpha _{g}})>
 \label{eq:5}
\end{equation}
\begin{equation}
 w(V,\bar{\alpha},\bar{\alpha _{g}})= \frac{ {\cal{M}}(V, \bar{\alpha })}
{{\cal{M}}(V, \bar{\alpha_ {g} })}  \label{eq:6}
\end{equation}
\begin{equation}
<w(V,\bar{\alpha},\bar{\alpha _{g}})> =\int_{ \Delta \Omega }
\int w(V,\bar{\alpha},\bar{\alpha _{g}})
\cdot \frac{g(V,\bar{\alpha _{g}} )
\cdot R(V, \Omega )}{D(\Delta \Omega ,\bar{\alpha _{g}})}
dVd\Omega  \label{eq:7}
\end{equation}
Then, the integral (\ref{eq:1}) can be approximated by using the set of the
$n_{\Delta \Omega}$ M.C. events  as:
\begin{equation}
I\simeq \frac{D(\Delta \Omega ,\bar{\alpha _{g}})
\cdot \sigma_{obs}(\bar{\alpha _{g}})}{n_{\Delta \Omega}}\cdot
\sum_{i=1}^{n_{\Delta \Omega}} w(V_{i},\bar{\alpha},\bar{\alpha _{g}})
\label{eq:8}
\end{equation}
Obviously the function $D(\Delta \Omega ,\bar{\alpha _{g}})$ can be
estimated as:
\begin{equation}
D(\Delta \Omega ,\bar{\alpha _{g}})\simeq \frac{n_{\Delta \Omega}}{N}
\label{eq:9}
\end{equation}
with a binomial error $\sqrt{n_{\Delta \Omega}\cdot (N-n_{\Delta \Omega})/N^{3}}$,
which for $N\gg n_{\Delta \Omega}$ becomes $\sqrt{n_{\Delta \Omega}}/N$.
The global term in (\ref{eq:8}) $\sigma_{obs}(\bar{\alpha _{g}})$
(independent of the particular $\Delta \Omega$
interval) , can be written as a fraction of
the total cross section $\sigma_{tot}(\bar{\alpha _{g}})$ in the form:
\begin{equation}
\sigma_{obs}(\bar{\alpha _{g}}) \simeq \sigma_{tot}(\bar{\alpha _{g}})
\cdot \frac{N}{N_{0}} \label{eq:10}
\end{equation}
with a fractional  error of $\sqrt{N\cdot (N_{0}-N))/N_{0}^{3}}$
which for $N_{0}\gg N$ becomes $\sqrt{N}/N_{0}$.
Finally by substituting (\ref{eq:9}) and (\ref{eq:10}) into ( \ref{eq:8})
the integral I is approximated as:
\begin{equation}
I\simeq \frac{ \sigma_{tot}(\bar{\alpha _{g}})}{N_{0}}
\sum_{i=1}^{n_{\Delta \Omega}} w(V_{i},\bar{\alpha},\bar{\alpha _{g}})
\label{eq:11}
\end{equation}
The accuracy  of this approximation can be quantified in terms of the variance
of $w(V_{i},\bar{\alpha},\bar{\alpha _{g}})$
and the statistical errors of  (\ref{eq:9}) and (\ref{eq:10}). Thus
the error in (\ref{eq:11}) is:
\begin{eqnarray}
\delta _{I} = \frac{ \sigma _{tot}(\bar{\alpha _{g}})}{N_{0}}
\cdot [<w(V,\bar{\alpha },\bar{\alpha _{g}})>^{2} \cdot
(n_{\Delta \Omega } + \frac{n_{\Delta \Omega }^{2}}{N}) + \nonumber \\
n_{\Delta \Omega }\cdot (<w^{2}(V,\bar{\alpha },\bar{\alpha _{g}})> -
<w(V,\bar{\alpha },\bar{\alpha _{g}})>^{2})]^{1/2}
\label{eq:12}
\end{eqnarray}

The above approximation (\ref{eq:11}) and its error estimation (\ref{eq:12})
are 
valid for every value of $\bar{\alpha}$ and $\bar{\alpha _{g}}$ as long as
the p.d.f. defined in (\ref{eq:ppp}) is not zero inside the
$\Delta \Omega$ interval.

\subsection{ The Mean Value of a Function of the Observed Kinematical
Vectors }

 It has been shown \cite{optim} that there are  functions of the
true kinematical vectors,
${\cal{O}}_{k}(V;\bar{\alpha _0})$ with $  k=1, \ldots ,\rho $
which locally (around $\bar{\alpha _0}$) carry the
whole  information concerning the $k^{th}$ parameter.
This is easily proven by expanding the p.d.f. in a Taylor
series around an initial set of values  $\bar{\alpha _{0}}$
of the parameters and keeping only the linear terms.
Furthermore the mean values of these functions, usually called Optimal
Observables, for any $\bar{\alpha }$ around
$\bar{\alpha _{0}}$,
can be expressed as  linear functions
of the parameters with known functions as slopes and intercepts.
The situation becomes more complicated when one has to take into account
detector effects. However, it can be shown \cite{tff} that in this case too,
there exist functions of the observed kinematical vectors which retain the same
information content as the Likelihood estimators.
In general \footnote{ In the following the index $k$ is dropped to simplify
the expressions.}
this function ${\cal{\omega}}(\Omega;\bar{\alpha _0})$ is defined as:
\begin{equation}
{\cal{\omega}}(\Omega;\bar{\alpha _0}) = \int {\cal{O}}(V;\bar{\alpha
_0})\cdot\frac{g(V,\bar{\alpha _{0}})
\cdot R(V,\Omega )}{\int g(V,\bar{\alpha _{0}})
\cdot R(V,\Omega )dV}dV
\label{eq:13}
\end{equation}

In other words,
the Optimal Observable in the most general case  is the mean value of
${\cal{O}}(V;\bar{\alpha _0})$ where the distribution of the vectors  V 's
corresponds to
$\bar{\alpha _{0}}$ parameter  values under the condition that the observed
kinematical vectors are equal to $\Omega$.
Equation (\ref{eq:13}) can be further simplified if expressed in terms of
${\cal{O}}(\Omega)$ as:
\begin{equation}
{\cal{\omega}}(\Omega;\bar{\alpha _0}) = {\cal{O}}(\Omega;\bar{\alpha _0})\cdot
r(\Omega ;\bar{\alpha _0})
\label{eq:14}
\end{equation}
where  $r(\Omega;\bar{\alpha _0} )$ is the  mean value of the
Optimal Observables
${\cal{O}}(V;\bar{\alpha _0})$ (as defined in (\ref{eq:13})) expressed in units
of ${\cal{O}}(\Omega;\bar{\alpha _0})$.
The function $r(\Omega ;\bar{\alpha _0})$ plays the
role of a correction function which has to be calculated from the M.C.
In practice  \cite{tff} (at least for the case of the TGC's) a very good
approximation is:
\begin{equation}
{\cal{\omega }}(\Omega ;\bar{\alpha _0}) \simeq {\cal{O}}(\Omega ;\bar{\alpha
_0})
\label{eq:15}
\end{equation}

The mean value of ${\cal{O}}(\Omega;\bar{\alpha _0})$ (in the following
Modified
Observable or M.O.)
\begin{equation}
<{\cal{O}}(\Omega ;\bar{\alpha _0})>_{\bar{\alpha }} = \int{\cal{O}}(\Omega
;\bar{\alpha _0})\int g(V,\bar{\alpha })
\cdot R(V,\Omega )dVd\Omega   \label{eq:17}
\end{equation}
in a region
around $\bar{\alpha _{0}}$ is, as before,
a linear function of the couplings $\bar{\alpha }$ but in
this case the slopes and intercepts are  convolutions of known physics
functions (matrix elements and phase space) and the resolution and efficiency
functions. Nevertheless it is simpler to calculate the mean value
$<{\cal{O}}(\Omega ;\bar{\alpha _0})>_{\bar{\alpha }}$ as a function of
$\bar{\alpha }$  (calibration curve)
by using  M.C.  events with
proper weights corresponding to the $\bar{\alpha }$ parameter values.
In parallel the set of the $N_{meas.}$
vectors, $\Omega _{meas.}$, accumulated in the real experiment is used to
measure the mean value of the Modified Observable as:
\begin{equation}
<{\cal{O}}(\Omega ;\bar{\alpha _{0}})>_{meas.} \simeq
\frac{1}{N_{meas}}\cdot\sum_{i=1}^{N_{meas.}}
{\cal{O}}(\Omega _{meas}^{i};\bar{\alpha _0})
\label{eq:16}
\end{equation}
Finally an estimation of the couplings is achieved by comparing the
experimental value (\ref{eq:16}) with the calibration curve \footnote{
It must be emphasized that although the method of defining the
calibration curve is valid for every $\bar{\alpha }$, the Modified Observable
is Optimal only in a narrow region around $\bar{\alpha }_{0}$. The estimation
technique
which extends the optimality to all the points of the parametric space
is discussed elsewhere \cite{tff}.}.

The reweighting procedure needed for the calculation of the calibration
curve  follows the same general principles as the method described in the
previous subsection.
As before, by multiplying and dividing the integrant of (\ref{eq:17}) by
$g(V,\bar{\alpha _{g} })$ ($\bar{\alpha _{g}}$ being the parameters used
for the production of the M.C. set of events) and expressing the p.d.f.'s
in terms of total cross sections,  squared matrix elements and phase
space factors, eq.  (\ref{eq:17}) is transformed to\footnote{Assuming that
$g(V,\bar{\alpha _{g} }) > 0 , \forall V$.}:
\begin{equation}
<{\cal{O}}(\Omega ;\bar{\alpha _0})>_{\bar{\alpha }} = \frac{\sigma
_{obs}(\bar{\alpha _{g} })}
{\sigma _{obs}(\bar{\alpha })}\cdot
\int\ \int {\cal{O}}(\Omega ;\bar{\alpha _0})\cdot w(V,\bar{\alpha
},\bar{\alpha } _{g})
\cdot g(V,\bar{\alpha_{g} })
\cdot R(V,\Omega )dVd\Omega   \label{eq:18}
\end{equation}
This integral can be approximated as:
\begin{equation}
<{\cal{O}}(\Omega ;\bar{\alpha _0})>_{\bar{\alpha }} \simeq
\frac{1}{\sum_{i=1}^{N} w(V_{i},\bar{\alpha},\bar{\alpha _{g}})}
\cdot  \sum_{i=1}^{N} w(V_{i},\bar{\alpha},\bar{\alpha _{g}})
\cdot {\cal{O}}(\Omega _{i};\bar{\alpha _0})
\label{eq:19}
\end{equation}
where the $\sigma _{obs}(\bar{\alpha })$ was expressed as
\begin{equation}
\sigma _{obs}(\bar{\alpha }) \simeq  \frac{\sigma _{obs}(\bar{\alpha _{g}})}{N}
\cdot \sum_{i=1}^{N} w(V_{i},\bar{\alpha},\bar{\alpha _{g}})
\label{eq:20}
\end{equation}

The variance of the approximation (\ref{eq:19}) can be estimated in the
standard way in terms of the variance, covariance and partial derivatives
of (\ref{eq:19}) with respect to the "random"
variables $w(V_{i},\bar{\alpha},\bar{\alpha _{g}})$ and
$w(V_{i},\bar{\alpha},\bar{\alpha _{g}})
\cdot {\cal{O}}(\Omega _{i};\bar{\alpha _0})$.

\section{Combining Several Sets of M.C. events}

 In practice, the use of a limited size M.C. set of events has the disadvantage
that although the p.d.f. could be  no zero, there are phase space regions where
very few (or none) events have been generated. The use of such a set to
estimate integrals or mean values, by reweighting, could induce systematical 
biases. These biases are important  when the particular not-covered phase 
space region contributes
significantly to the cross section corresponding to the extrapolated
parameter values. Thus the use of several M.C. sets produced at different
points of the parametric space and properly combined together is preferable.

Let ${\cal{Q}}(\bar{\alpha })$ be the integral of a function of the observed
kinematical vectors (e.g. cross section or mean value).
Let  $S_{j}$  $(m = 1, \ldots ,m )$  be m sets of M.C. events produced with
parameter values $ \bar{\alpha _{1}} , \ldots , \bar{\alpha _{m}}$. Let then
$\hat{{\cal{Q}}}_{i}(\bar{\alpha })$ be an estimation of
${\cal{Q}}(\bar{\alpha })$ with variance ${\cal{V}}_{i}(\bar{\alpha })$
where the $i^{th}$ set of M.C. events has been used as it is
described in the previous section. Then a better estimation of
${\cal{Q}}(\bar{\alpha })$ can be found as a linear combination
of the $\hat{{\cal{Q}}}_{i}(\bar{\alpha })$ 's, i.e.
\begin{equation}
\hat{{\cal{Q}}}_{comb.}(\bar{\alpha }) =
\sum_{i=1}^{m} \gamma _{i}\cdot \hat{{\cal{Q}}}_{i}(\bar{\alpha })
\end{equation}
\begin{equation}
\sum_{i=1}^{m} \gamma _{i} =1
\label{eq:21}
\end{equation}
such as the variance of $\hat{{\cal{Q}}}_{comb.}(\bar{\alpha })$ to be
minimum. It is easy to show that this is equivalent to the minimization of 
the least square sum 
\begin{equation}
{\cal \chi}^{2} =
\sum_{i=1}^{m} (\hat{{\cal{Q}}}_{comb.}(\bar{\alpha })- \hat{{\cal{Q}}}_{i}(\bar{\alpha }))^{2}
/{{\cal{V}}_{i}(\bar{\alpha })}
\label{eq:21a}
\end{equation}
which results to:
\begin{equation}
\hat{{\cal{Q}}}_{comb.}(\bar{\alpha }) =
\sum_{i=1}^{m} (\frac{\hat{{\cal{Q}}}_{i}(\bar{\alpha })}
{{\cal{V}}_{i}(\bar{\alpha })}) / (\sum_{i=1}^{m}\frac{1}
{{\cal{V}}_{i}(\bar{\alpha })})
\label{eq:22}
\end{equation}
with a variance of:
\begin{equation}
V(\hat{{\cal{Q}}}_{comb.}(\bar{\alpha })) =
1./ (\sum_{i=1}^{m}\frac{1}
{{\cal{V}}_{i}(\bar{\alpha })})
\label{eq:22a}
\end{equation}

There are however cases, where integrals of more than one functions have to be estimated 
(e.g. in the case of two-T.G.C. simultaneous estimation, two M.O. have to be 
evaluated as functions of the two couplings) 
by integration using the same sets of M.C. events. The generalization of (\ref{eq:21a})
for n integrals reads:
\begin{equation}
{\cal \chi}^{2} = \sum_{i=1}^{m}
(\hat{{\cal\bf{Q}}}_{comb.}(\bar{\alpha })- \hat{{\bf{Q}}}_{i}(\bar{\alpha }))^{T}
\cdot {\bf M}^{-1}_{i} \cdot
(\hat{{\cal\bf{Q}}}_{comb.}(\bar{\alpha })- \hat{{\bf{Q}}}_{i}(\bar{\alpha }))
\label{eq:22b}
\end{equation}
where the vectors 
$\hat{{\cal\bf{Q}}}_{comb.}(\bar{\alpha })$ and $\hat{{\bf{Q}}}_{i}(\bar{\alpha })$
have elements coresponding to each of the n  integrated functions, whilst
the subscript $i$ stands for the set of the M.C. events used in the integration.
The matrices ${\bf M}_{i}$ represent the covariant matrices between the elements of 
the vector $\hat{{\bf{Q}}}_{i}(\bar{\alpha })$.

\section{Numerical Results}

For any practical application of the reweighting technique,
one needs only the means to calculate
the weight $ w(V,\bar{\alpha},\bar{\alpha _{g}})$ as it is defined in (\ref{eq:6}).
Concentrating on the physics analysis concerning the determination of
the TGC's there are several four
fermion
M.C. generators which include the relevant physics processes \cite{yb}. Among
them, the ERATO
generator \cite{erato} has been chosen and was modified keeping only the 
parts necessary for
the calculation of the squared matrix element (M.E.). In case that an ISR
photon
(with or without $P_{T}$) had been emitted, the four fermions kinematical vectors were
transformed to their rest frame before calling 
the ERATO M.E. calculating routines. Here the ISR effect 
is assumed to be
factorizable and thus independent of the variables. Thus,
as it has been shown previously with 
the  phase space factors, the ISR factors in the weight definitions
are canceled out.

In general lines one should a) define the values of the physics constants 
to be used for the matrix element calculation,
b) pass the kinematical
and event type information and c) call the relevant software twice
to calculate the ratio of the M.E.'s as in (\ref{eq:6}).

In the following, the reweighting method is demonstrated through applications 
concerning the evaluation of
physical quanitites needed in the estimation of the TGC's, 
using four fermion (specifically the lepton-neutrino-two hadronic jets)
final state events as those produced at LEPII at 172 GeV 
centre of mass energy. 

The available M.C. events undergone full detector simulation 
(DELSIM \cite{delsim}) are grouped into two main categories.
\begin{itemize} 
\item Events produced with PYTHIA \cite{pythia} with the TGC's set to their Standard 
Model (S.M.) values whilst six different W mass (79.35,79.85,80.23,80.35,80.85 and 81.35 GeV/$c^{2}$)
values were used. These 
six subcategories contain in all  3280 and 3390 events having an electron 
(electron type) or a muon (muon type) in the final state respectively. 
\item Events produced with EXCALIBUR \cite{exca} with the W mass set to 80.35 
GeV/$c^{2}$ but divided into three groups according to the $\alpha_{ W\phi} $ values 
used in the production whilst the other couplings are related according to the 
$W\phi $ model \cite{yb}. Namely
the $\alpha _{W\phi } = 0$ subgroup contains 790 electron type and 900 muon type 
events,
the $\alpha _{W\phi } = -2$ subgroup contains 770 electron type and 900 muon type 
events and
the $\alpha _{W\phi } = 2$ subgroup contains 810 electron type and 900 muon type 
events.
\end{itemize}

Identical selection criteria have beee applied on both generation and reconstruction level
to all the M.C. set of events.
The PYTHIA physics generator employs only the resonant Feynman graphs (CC03) whilst the 
EXCALIBUR  generator includes  all the 
four fermion (CC10 for muon type and CC20 for electron type) relevant processes. For all the applications
to follow the extrapolated parameter variables correspond to the full four fermion
production mechanisms and the extrapolated W mass value was  equal to 80.35 GeV/$c^{2}$ 
whilst the 
estimated quantities are considered as functions of the $\alpha_{W\phi}$ coupling.

 The simplest demonstration of the reweighting technique
deals with the estimation of the total cross section as a function of the coupling. This in accordance
to (\ref{eq:11}) can be estimated by using the $j^{th}$ set of the available M.C. events as: 
\begin{equation}
I_{j} \simeq \frac{ \sigma_{tot}( \bar{ \alpha}_{ W\phi} ^{j} )}{N_{0}^{j}}\cdot
\sum_{i=1}^{N_{0}^{j}} w(V_{i},\bar{\alpha_{ W\phi}},\bar{\alpha}_{  W\phi}^{j}) 
\label{eq:31}
\end{equation}
with a statistical error equal to:

\begin{equation}
\delta _{j} = \frac{ \sigma _{tot}(\bar{\alpha}_{ W\phi}^{j} )}{\sqrt{N_{0}}}  
\cdot [ 
 <w^{2}(V,\bar{\alpha_{ W\phi}},\bar{\alpha}_{ W\phi}^{j} )> -
<w(V,\bar{\alpha_{ W\phi}},\bar{\alpha}_{ W\phi}^{j})>^{2}]^{1/2} 
\label{eq:32}
\end{equation}

In the following  several M.C. set of events are combined as in (\ref{eq:22}) 
to increase the statistical accuracy
of this approximation. Figure \ref{cross}a and \ref{cross}b presents the estimated 
dependence of the number of  produced events for a 9.6 $pb^{-1}$ accumulated luminosity 
on the value of $\alpha_{ W\phi}$ for electronic and muonic final states when only 
the PYTHIA M.C. sets have been reweighted. 
This estimation of the reweighting procedure agrees,
for coupling values between -2 and 2,
 within statistical errors
with the  prediction (solid line) of 
the ERATO generator containing the full set of four fermion production diagrams. 
However, this extrapolation 
suffers from the fact that the PYTHIA generated events do not 
cover adequately all the phase space regions which are contributing significantly to the
four fermion cross section for larger absolute values of the couplings.
An enlargement of the extrapolation region with a simultaneous improvement of the 
accuracy is achieved when the  EXCALIBUR M.C. sets are also included as it can 
be seen  in figures  \ref{cross}c and \ref{cross}d.

The second example is dealing with integrations with respect to the observed 
kinematical vectors.
In this case the efficiency is estimated as the fraction of the number
of the produced events (eq. \ref{eq:31}), which are selected in the final 
analysis sample (eq. \ref{eq:11}).   
As it is shown  in fig. \ref{effic} the estimated efficiency by reweighting is in a 
very good agreement
with the results of direct calculations  using  the EXCALIBUR M.C. sets 
produced at different values of $\alpha_{ W\phi}$.

The success of the reweighting technique to perform integrations with respect 
to the observed kinematical vectors can be also demonstrated by
comparing  estimations of  the number of events to be observed in a special 
phase space region, by reweighting and by direct M.C. integration.
In this example only the PYTHIA M.C. sets were used to estimate  these
differential quantities as functions of $\alpha_{ W\phi}$ by employing reweighting. 
In parallel the EXCALIBUR M.C. sets were used for a direct estimation at the three
available values of $\alpha_{ W\phi}$. These integrals represent the number of the 
expected
events for a total luminosity of 9.6 $pb^{-1}$, in bins either of $\cos{\theta_W}$ 
(Fig \ref{cwwe0} to Fig \ref{cwwep2})
($\theta_{W}$ being the polar angle of the hadronic system) or of the Modified 
Observable (Fig \ref{opte0} to Fig \ref{optep2}).
As a measure of the level of agreement in this comparison, the relative difference 
of the two  estimations was chosen (relative deviation) but the estimated numbers of 
events are also presented as insets in the previous figures. The remarkable agreement 
between the two type of estimations justifies the use of the reweighting technique 
in producing the distribution 
shapes for a wide area of the  $\alpha_{ W\phi}$ to be used in fitting the TGC's.

As extensively discussed previously, it is possible to define a Modified Observable
such as  its mean value is an estimator of the $\alpha_{ W\phi}$ which  carries the 
same information as the likelihood. It has been also
shown that (independent of the optimality) the evolution of this
quantity as a function of the coupling can be accurately approximated
by the reweighting procedure. Indeed, as it is shown in Fig. \ref{optimean}, 
the mean value of the Modified 
Observables defined for $\alpha_{ W\phi} = 0$ and estimated by reweighting is in 
a very good agreement with  the direct estimations by using the  EXCALIBUR M.C. sets. 
However  the optimality of this method is restricted to the linear
part around $\alpha_{ W\phi} = 0$. The extention of this optimality is achieved 
by expanding the p.d.f. in a Taylor series around another value of the 
coupling and reformulating the Modified Observables.
This procedure for optimal definition
has been employed  around the two extreme values of $\alpha_{W\phi} =  \pm 2$.
The results  are shown in Fig \ref{optimat} where also for comparison the
directly estimated values with the  EXCALIBUR M.C. sets are presented.

\section{Conclusions}
Following very simple calculus, it has been shown that a M.C. set of events
generated at specific values of the relevant physical parameters
can be used to approximate physical quantities in a very wide region
of the parametric space. Furthermore, a simple procedure has been demonstrated
which combines several M.C. samples, independent of the specific values of the
physical parameters at their generation. The quantification of the accuracy of this
approximation has been expressed analytically and demonstrated by numerical examples.
The applicability of this technique to produce the shapes of kinematical quantities
to be used in extracting the values of physical parameters was proven.

\topmargin=-2truecm
\textheight=26cm
\clearpage
 
\begin{figure}[cross]
\centerline{\epsfig{file=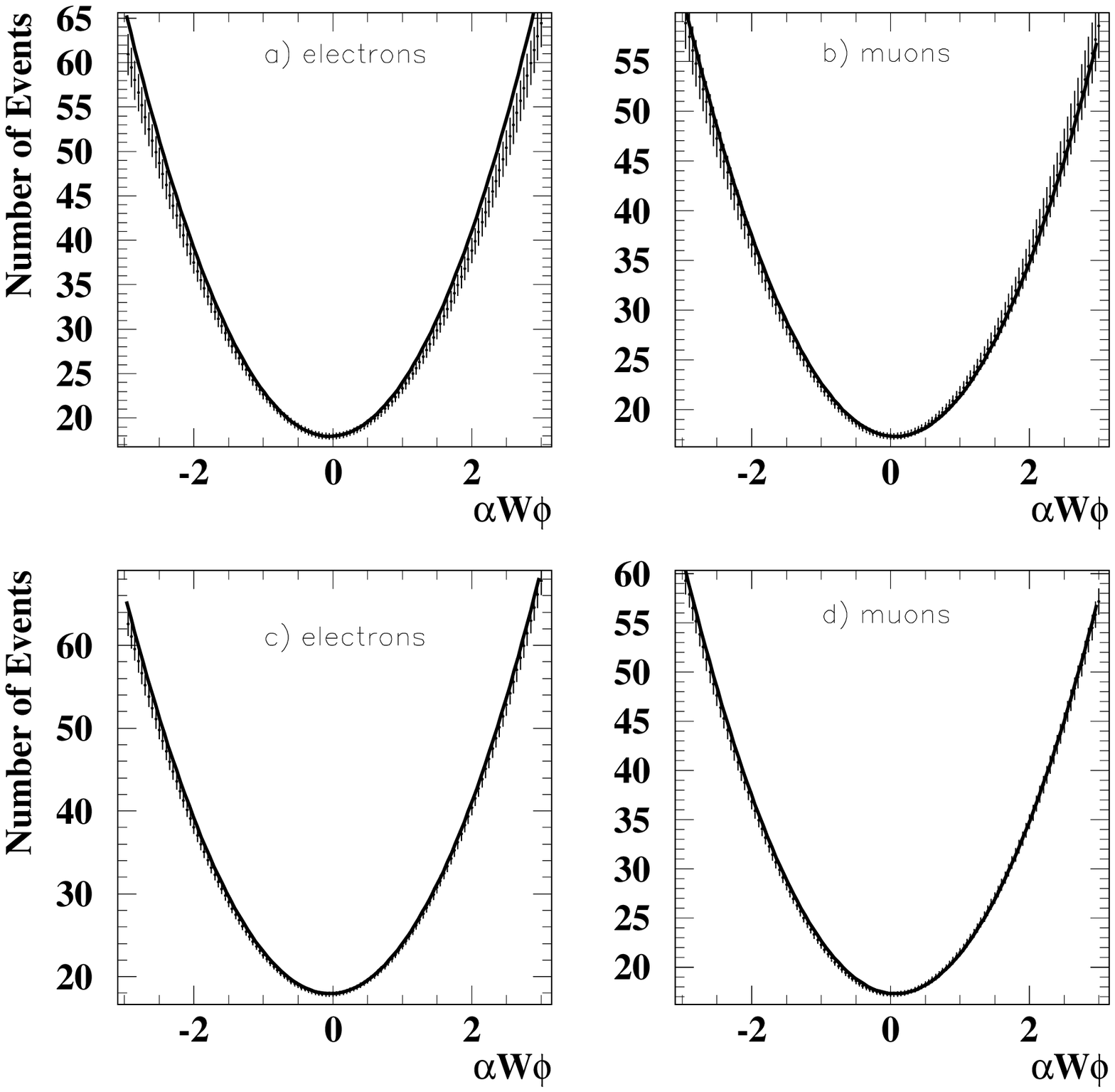,height=22cm}}
\caption{The total Cross Section as a function of the $\alpha_{ W\phi}$:
a) and b) only the PYTHIA,
c) and d) the PYTHIA and the EXCALIBUR Monte Carlo sets are  included in the 
reweighting. }
{
\label{cross}}
\end{figure}

\clearpage

\begin{figure}[effic]
\centerline{\epsfig{file=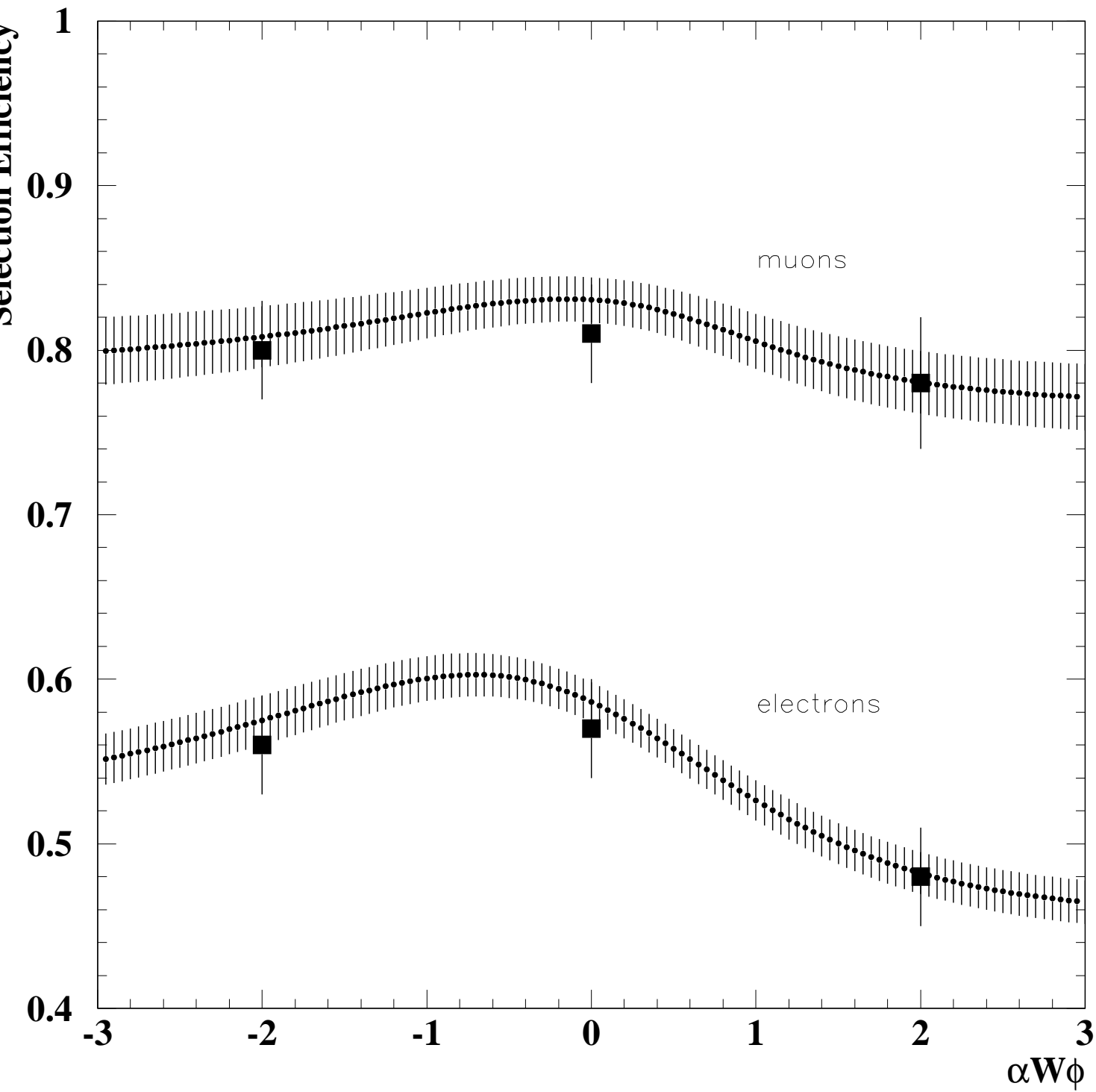,height=22cm}}
\caption{The Efficiency as a function of the $\alpha_{ W\phi}$ estimated
by reweighting only the PYTHIA Monte Carlo sets. The squares represent
the efficiency values estimated directly from the EXCALIBUR samples.}
{
\label{effic}}
\end{figure}

\clearpage

\begin{figure}[cwwe0]
\centerline{\epsfig{file=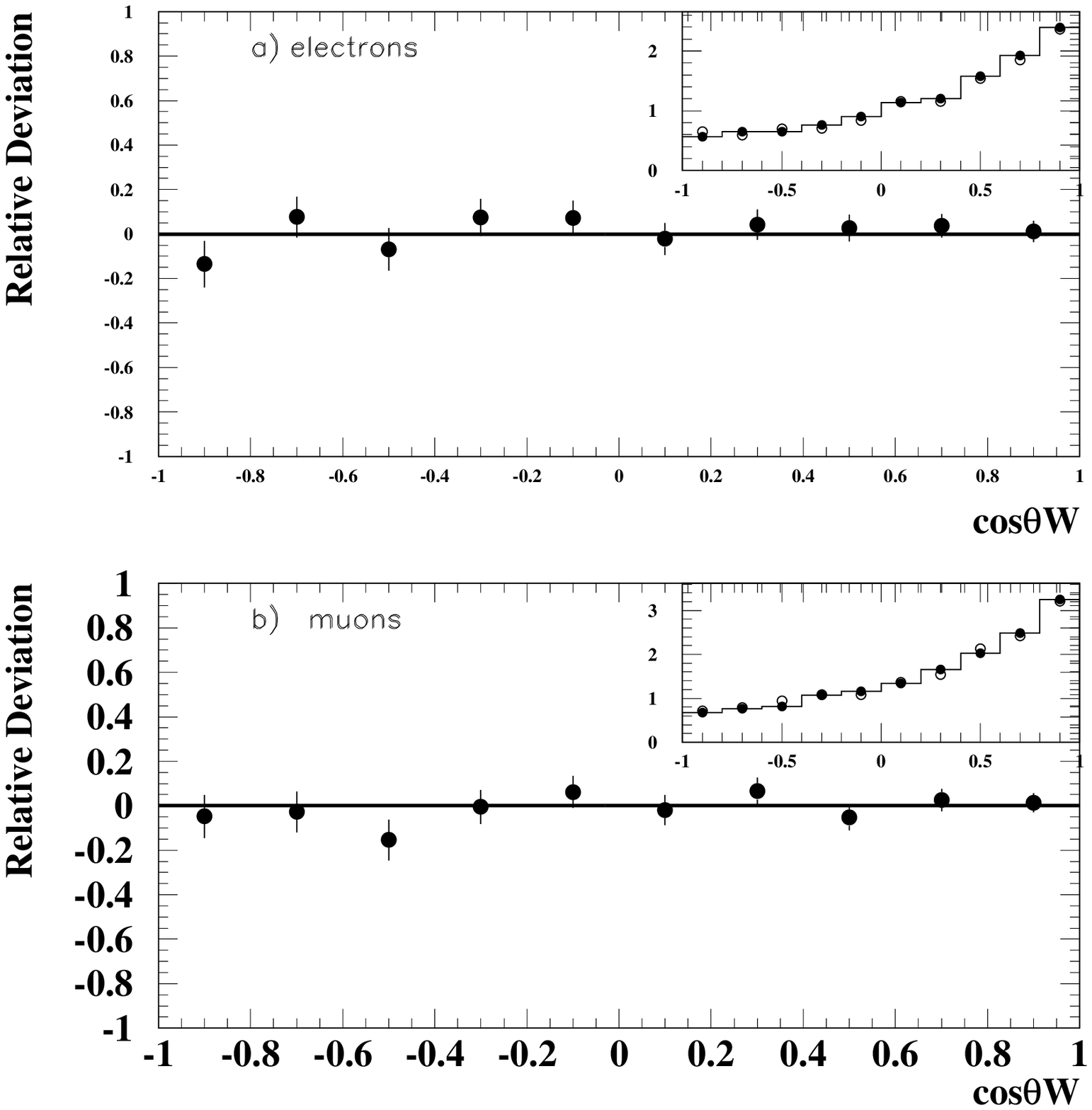,height=22cm}}
\caption{Comparison of the differential cross section
with respect to the $\cos{\theta _W}$ as it is estimated by
reweighting the PYTHIA sets  with  the distribution of unweighted
Monte Carlo events produced with EXCALIBUR at $\alpha_{ W \phi} = 0$.  
The solid lines and the black points in the inset figures correspond to the reweighting
estimation whilst the open circles stand for the EXCALIBUR prediction. }
{
\label{cwwe0}}
\end{figure}

\begin{figure}[cwwem2]
\centerline{\epsfig{file=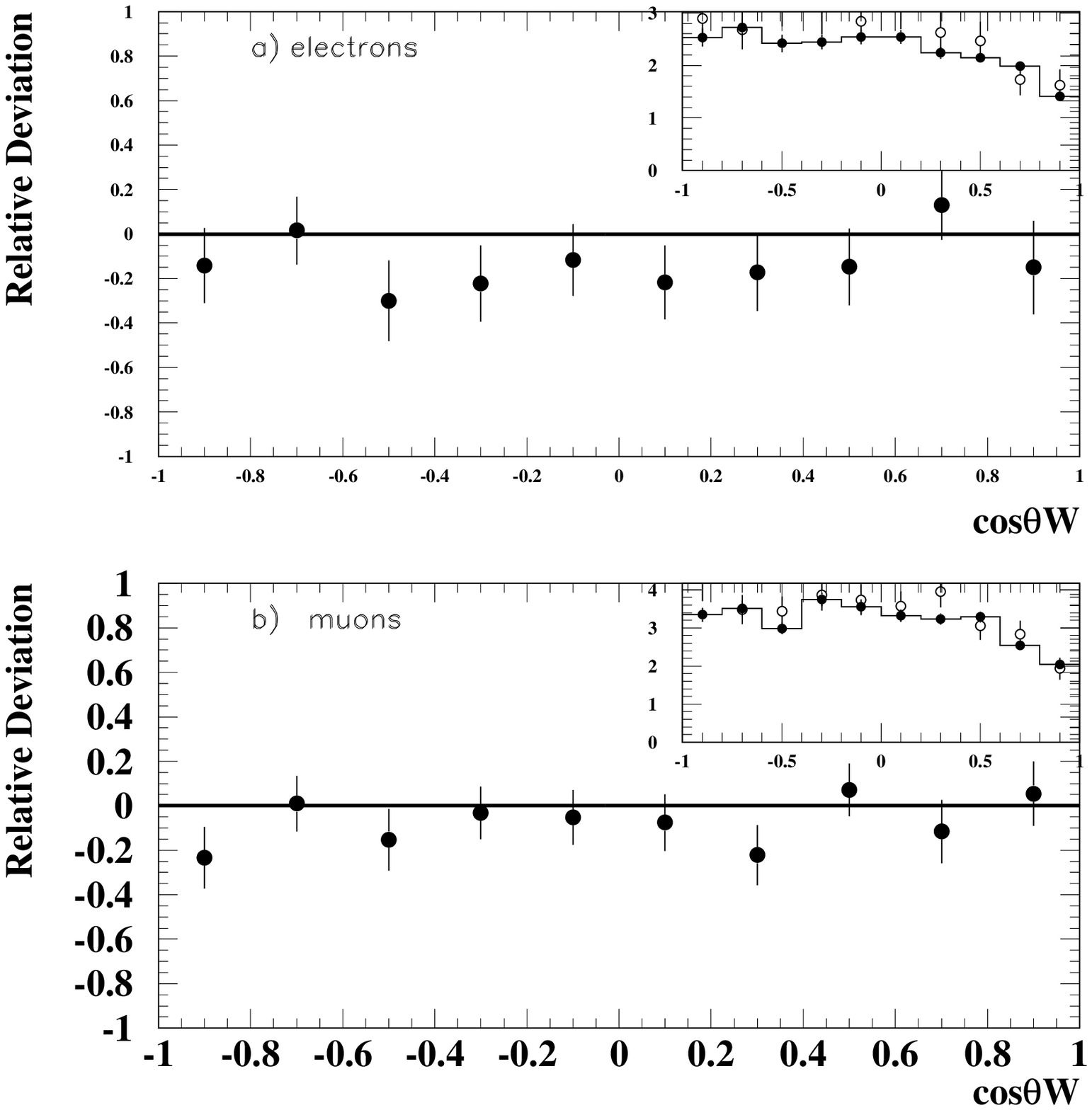,height=22cm}}
\caption{Comparison of the differential cross section
with respect to the $\cos{\theta _W}$ as it is estimated by
reweighting the PYTHIA sets  with  the distribution of unweighted
Monte Carlo events produced with EXCALIBUR at $\alpha_{ W \phi} = -2$.
The solid lines and the black points in the inset figures correspond to the reweighting
estimation whilst the open circles stand for the EXCALIBUR prediction.  }
{
\label{cwwem2}}
\end{figure}

\begin{figure}[cwwep2]
\centerline{\epsfig{file=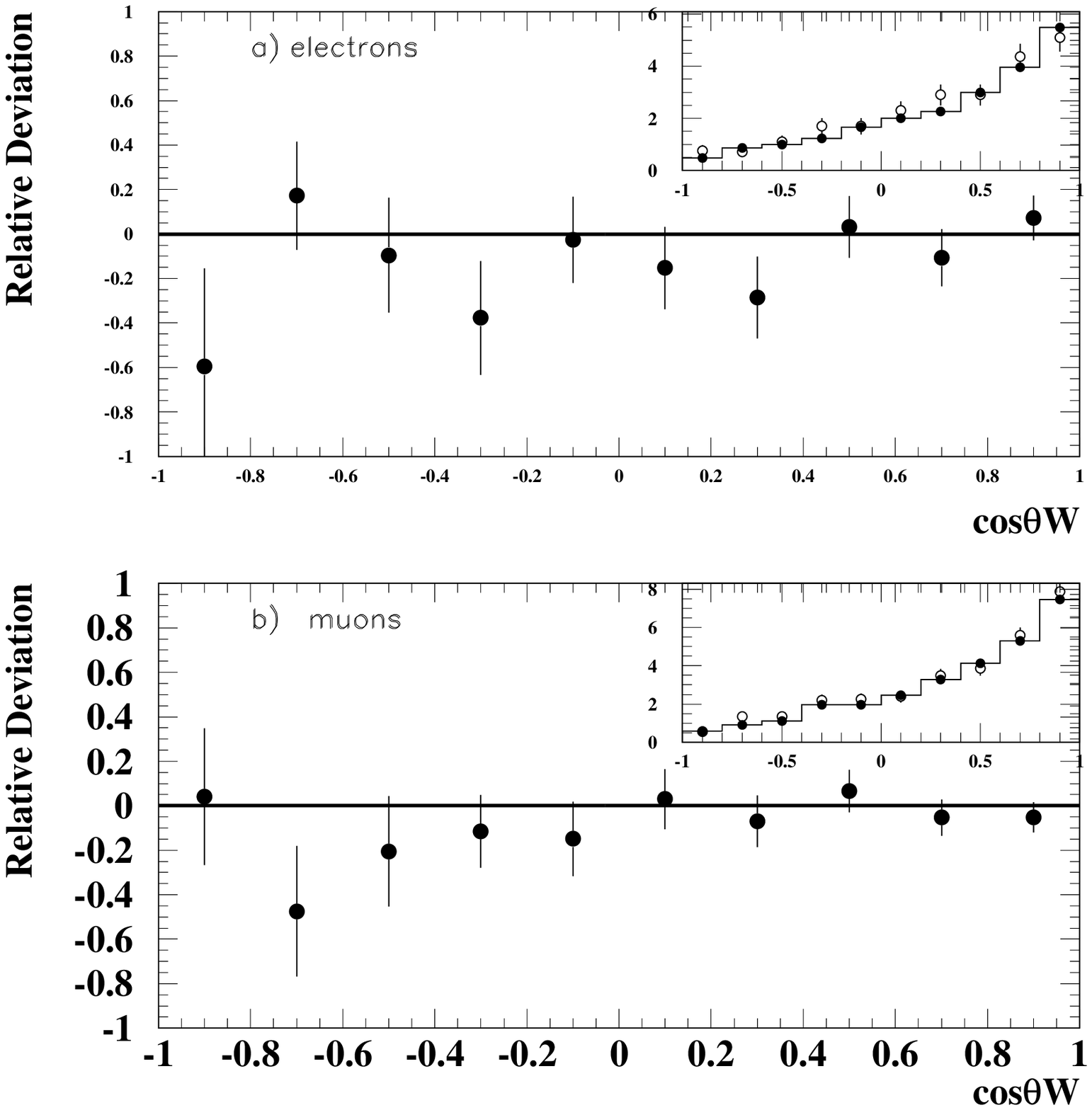,height=22cm}}
\caption{Comparison of the differential cross section
with respect to the $\cos{\theta _W}$ as it is estimated by
reweighting the PYTHIA sets  with  the distribution of unweighted
Monte Carlo events produced with EXCALIBUR at $\alpha_{ W \phi} = +2$.
The solid lines and the black points in the inset figures correspond to the reweighting
estimation whilst the open circles stand for the EXCALIBUR prediction.  }
{
\label{cwwep2}}
\end{figure}

\begin{figure}[opte0]
\centerline{\epsfig{file=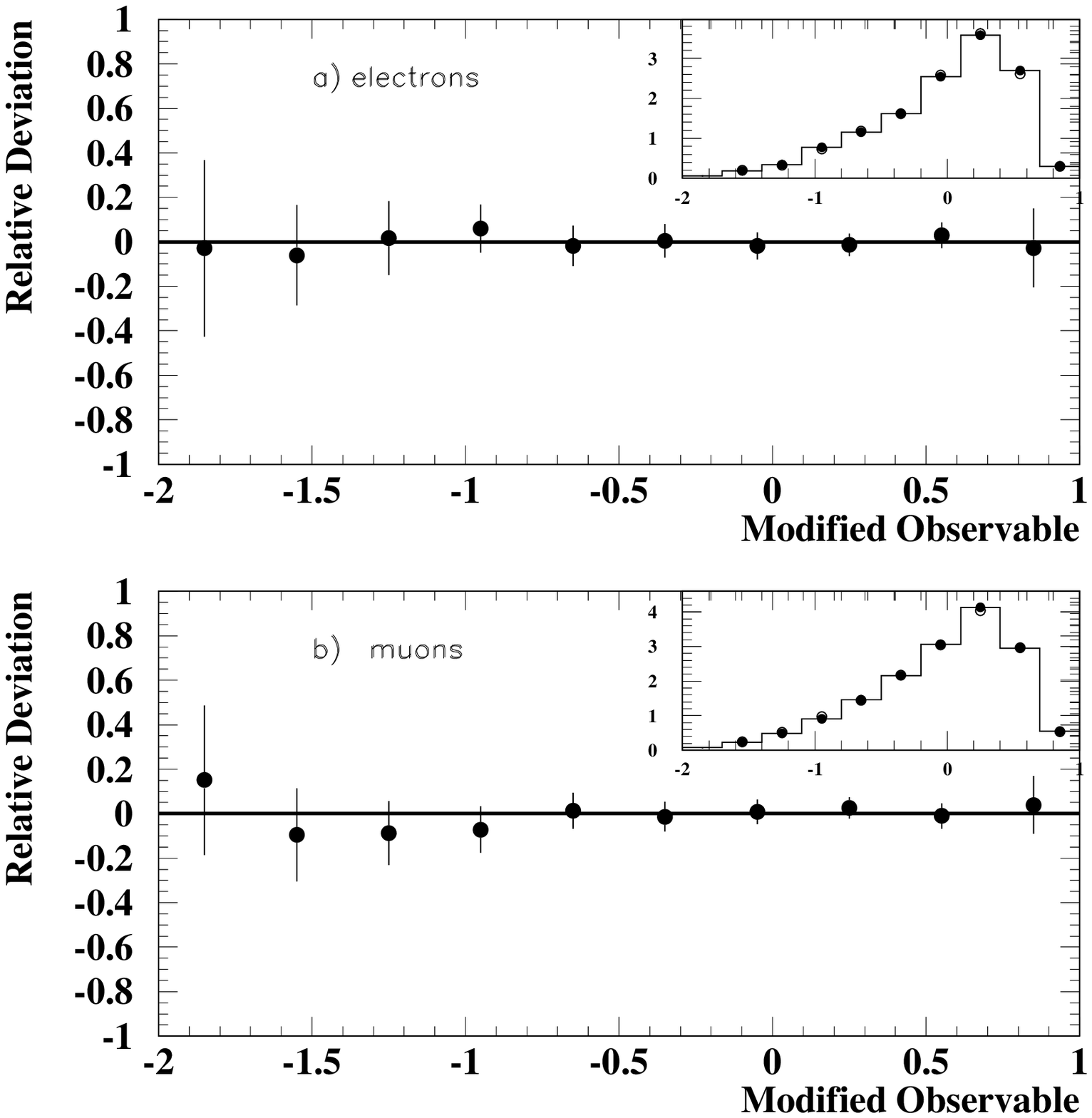,height=22cm}}
\caption{Comparison of the differential cross section
with respect to the Modified Observables (defined at $\alpha_{ W \phi} = 0$)
as it is estimated by
reweighting the PYTHIA sets  with  the distribution of unweighted
Monte Carlo events produced with EXCALIBUR at $\alpha_{ W \phi} = 0$.
The solid lines and the black points in the inset figures correspond to the reweighting
estimation whilst the open circles stand for the EXCALIBUR prediction.  }
{
\label{opte0}}
\end{figure}

\begin{figure}[optem2]
\centerline{\epsfig{file=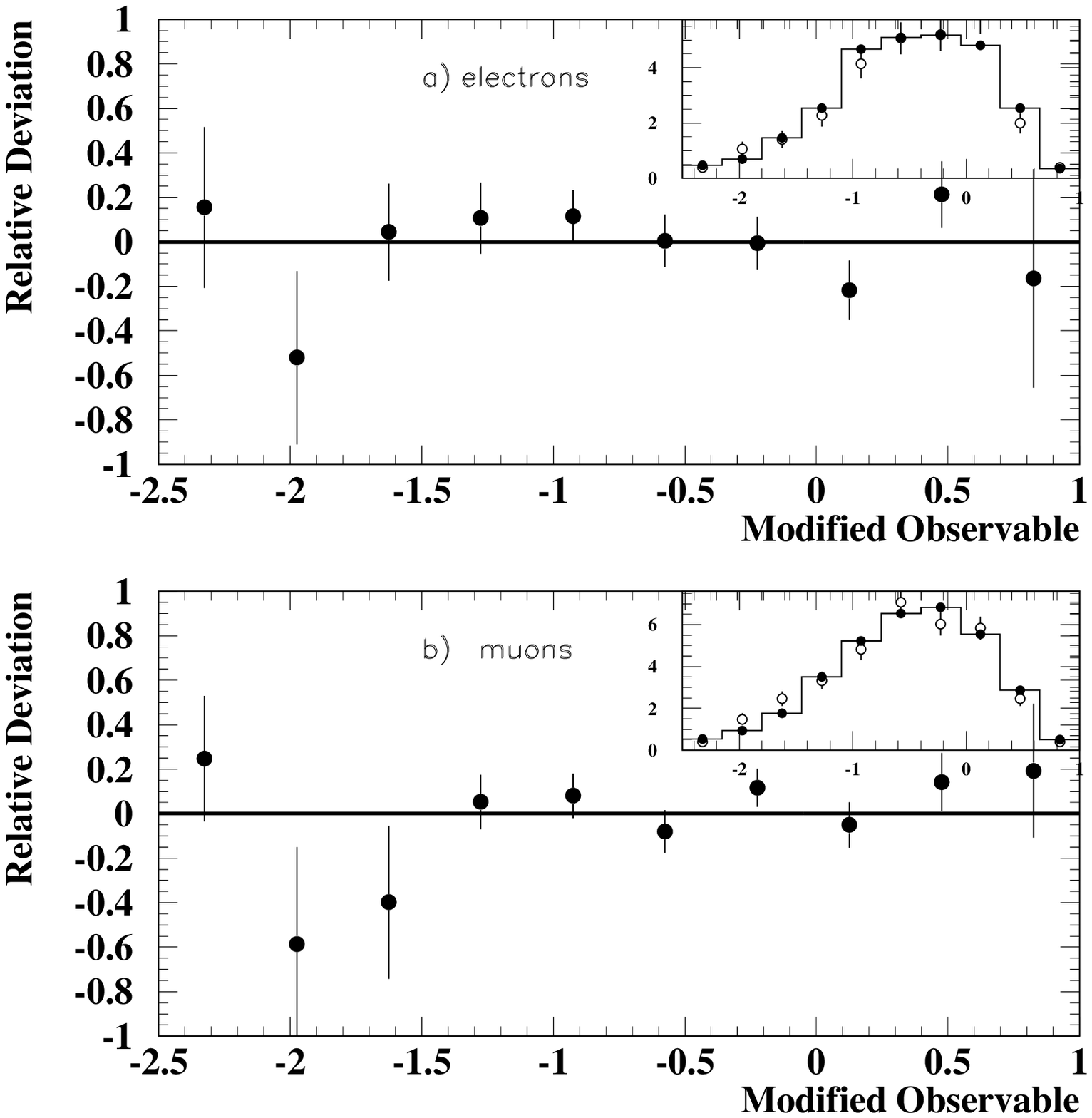,height=22cm}}
\caption{Comparison of the differential cross section
with respect to the Modified Observables (defined at $\alpha_{ W \phi} = 0$)
as it is estimated by
reweighting the PYTHIA sets  with  the distribution of unweighted
Monte Carlo events produced with EXCALIBUR at $\alpha_{ W \phi} = -2$.
The solid lines and the black points in the inset figures correspond to the reweighting
estimation whilst the open circles stand for the EXCALIBUR prediction.  }
{
\label{optem2}}
\end{figure}

\begin{figure}[optep2]
\centerline{\epsfig{file=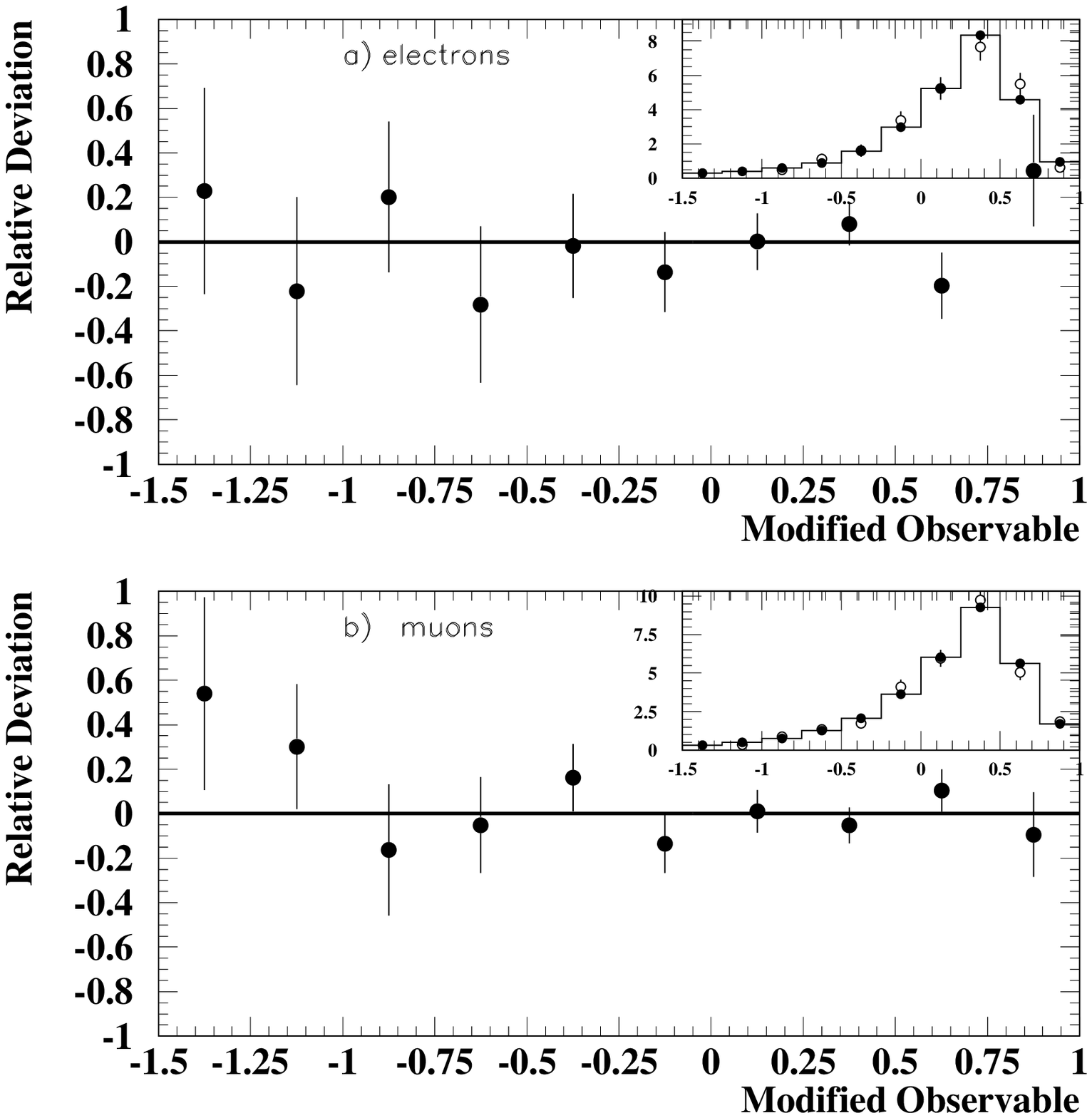,height=22cm}}
\caption{Comparison of the differential cross section
with respect to the Modified Observables (defined at $\alpha_{ W \phi} = 0$)
as it is estimated by
reweighting the PYTHIA sets  with  the distribution of unweighted
Monte Carlo events produced with EXCALIBUR at $\alpha_{ W \phi} = +2$.
The solid lines and the black points in the inset figures correspond to the reweighting
estimation whilst the open circles stand for the EXCALIBUR prediction.  }
{
\label{optep2}}
\end{figure}

\begin{figure}[optimean]
\centerline{\epsfig{file=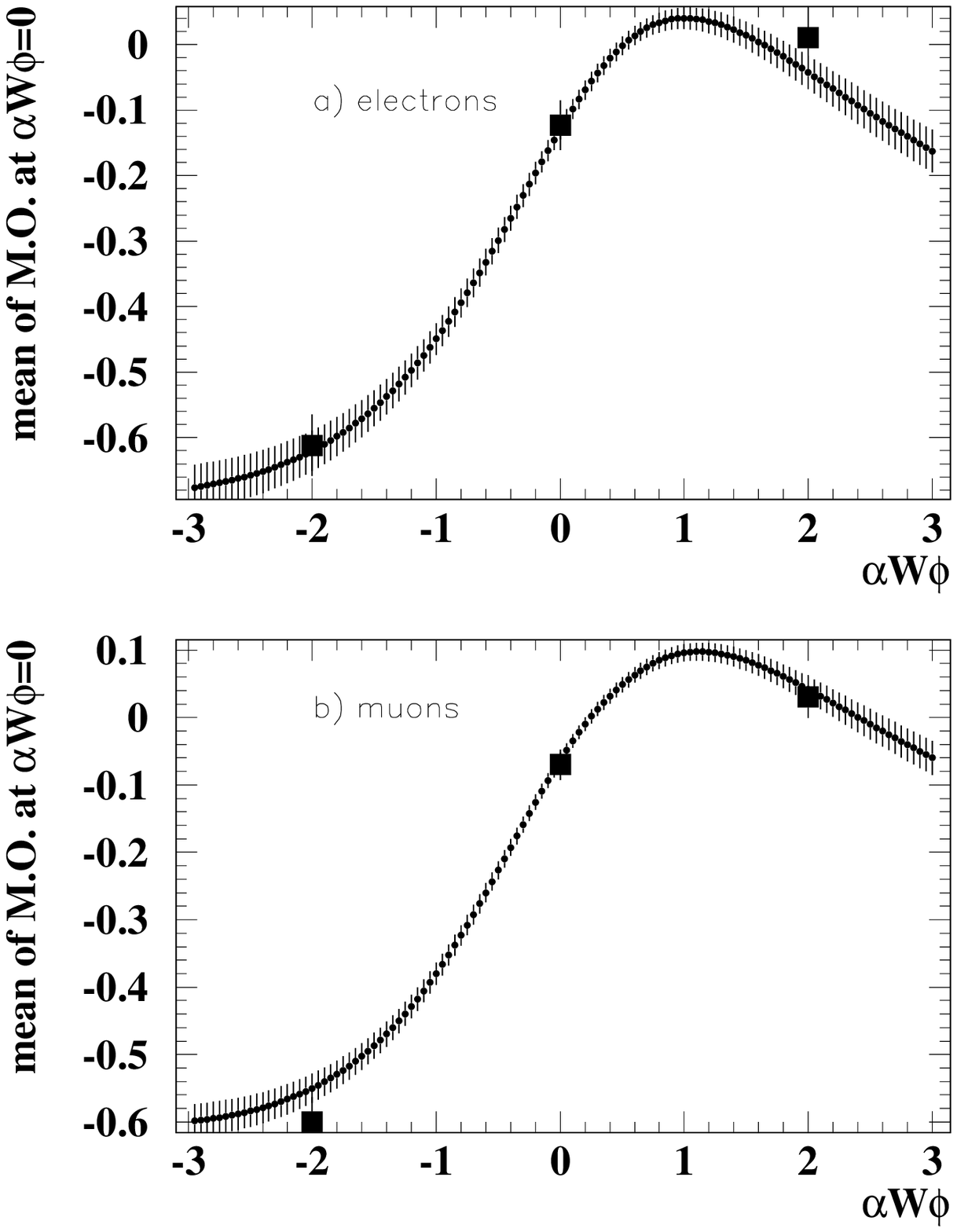,height=22cm}}
\caption{The Mean Value of the Modified Observable
 defined at $\alpha_{ W\phi} = 0$, as a function of the $\alpha_{ W\phi}$.
The squares correspond to unweighted EXCALIBUR events.  }
{
\label{optimean}}
\end{figure}

\begin{figure}[optimat]
\centerline{\epsfig{file=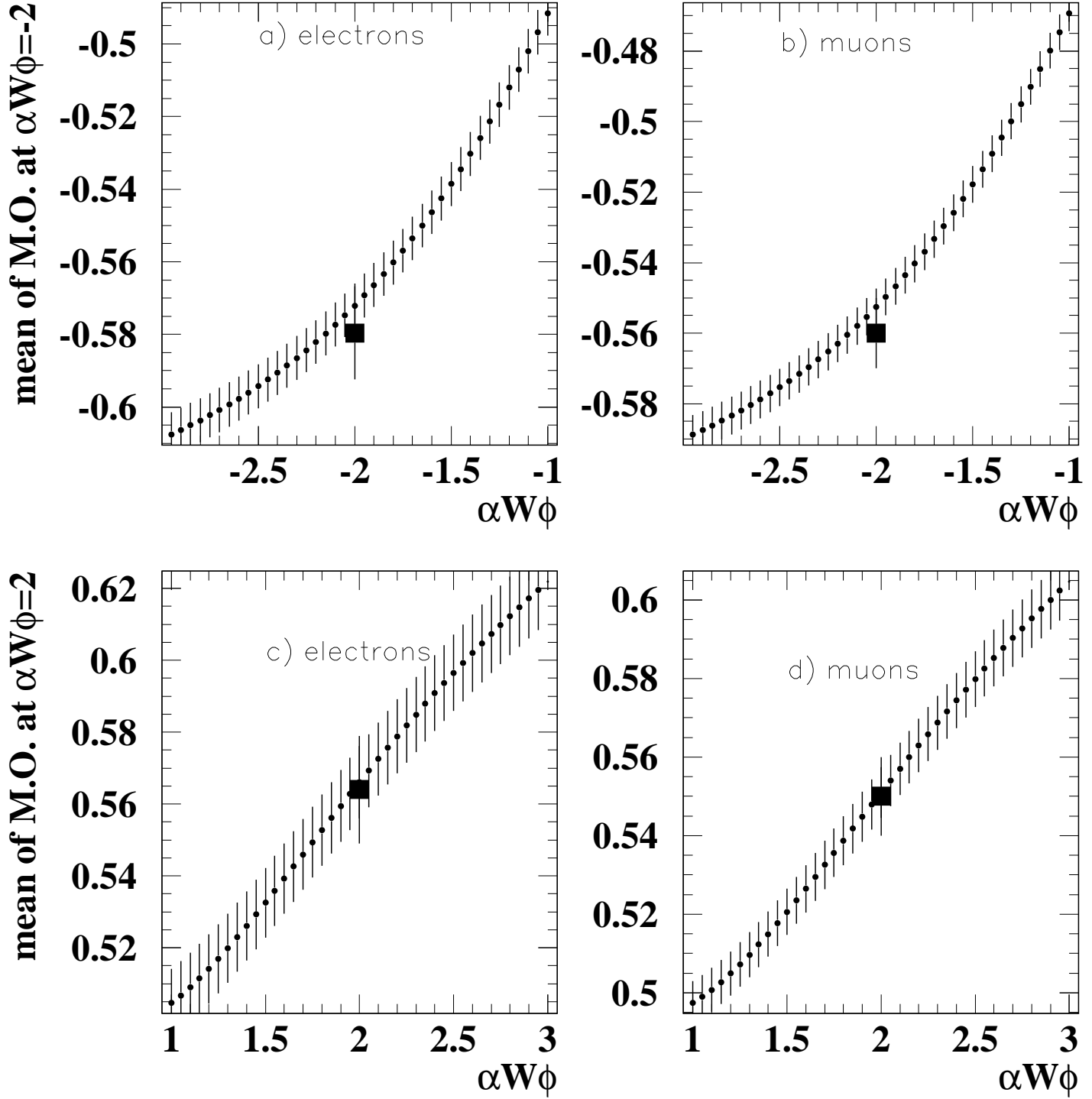,height=22cm}}
\caption{The Mean Value of the Modified Observable defined at
 (a and b) $\alpha_{ W\phi}$ equal to -2 and (c and d) at 2 
 as a function of the $\alpha_{ W\phi}$.
The squares correspond to unweighted EXCALIBUR events. }
{
\label{optimat}}
\end{figure}

\vfill
\end{document}